\documentclass[fleqn,10pt]{SelfArx} 

\definecolor{color1}{RGB}{0,0,90} 
\definecolor{color2}{RGB}{0,20,20} 

\usepackage{hyperref} 
\hypersetup{hidelinks,colorlinks,breaklinks=true,urlcolor=color2,citecolor=color1,linkcolor=color1,bookmarksopen=false,pdftitle={Title},pdfauthor={Author},urlcolor=blue}

\usepackage{graphicx}
\usepackage[square]{natbib}
\usepackage{amsmath}
\usepackage{multirow}
\usepackage{subcaption}
\usepackage{pdflscape}
\usepackage[final]{pdfpages}

\usepackage{graphicx}
\usepackage{natbib}
\usepackage{amsmath}
\usepackage{amsfonts}
\usepackage{multirow}
\usepackage{tikz}
\usetikzlibrary{calc}

\DeclareMathOperator{\sgn}{sgn}

\DeclareMathOperator{\atantwo}{atan2}

\usepackage{mathtools,amssymb,lipsum}

\usepackage{cuted}
\usepackage{boondox-calo}

\usepackage{calligra}

\DeclareMathAlphabet{\mathcalligra}{T1}{calligra}{m}{n} \DeclareFontShape{T1}{calligra}{m}{n}{<->s*[2.2]callig15}{}

\newcommand{\ud}{\mathrm{d}}
\newcommand{\xhat}{\mathbf{\hat{x}}}
\newcommand{\yhat}{\mathbf{\hat{y}}}
\newcommand{\zhat}{\mathbf{\hat{z}}}
\newcommand{\nhat}{\mathbf{\hat{n}}}
\newcommand{\rhat}{\mathbf{\hat{r}}}

\newcommand{\xr}{r}
\newcommand{\Dz}{\Delta z}
\newcommand{\bfr}{\mathbf{r}}
\newcommand{\bfM}{\mathbf{M}}
\newcommand{\bfH}{\mathbf{H}}

\newcommand{\bfN}{\mathbb{N}}
\newcommand{\hA}{\mathbb{A}}
\newcommand{\hC}{\mathbb{C}}
\newcommand{\hD}{\mathbb{D}}
\newcommand{\hE}{\mathbb{E}}
\newcommand{\bx}[5]{\Bigl[\,#1\,\Bigr]_{#2,\,#3}^{#4,\,#5}}
\newcommand{\bxs}[1]{\Bigl[\,#1\,\Bigr]}

\JournalInfo{~} 
\Archive{~} 

\PaperTitle{The magnetic field of a homogeneously magnetized circular piece} 

\Authors{Kaspar K. Nielsen, Rasmus Bjørk} 
\affiliation{\textit{Department of Energy Conversion and Storage, Technical University of Denmark - DTU, DK-2800 Kgs. Lyngby, Denmark}} 
\affiliation{*\textbf{Corresponding author}: rabj@dtu.dk} 

\newcommand{\keywordname}{Keywords} 

\Abstract{The magnetic field of a homogeneously magnetized circular piece, i.e.\ the part of a finite cylinder that is cut off by two planes parallel to the coordinate axes, is determined. The field is expressed as the product between the demagnetization tensor describing the geometrical part of the problem and the magnetization. All nine components of the tensor are expressed in closed form in terms of elementary functions and incomplete elliptic integrals of the first, second and third kind. The complementary geometrical piece, bounded by the same arc from the outside, is obtained from the same expressions. The solution, implemented in the open source magnetic framework MagTense, is verified against direct numerical integration and against a finite element calculation of the magnetic field, showing excellent agreement.}

\begin{document}

\flushbottom 

\maketitle 


\thispagestyle{empty} 

\section{Introduction}
\label{sec:intro}
Magnetic fields are used in a large variety of applications such as permanent magnet motors and generators \citep{Zhu_1993,Markovic_2009}, magnetic gears and bearings \citep{Lubin_2010,Ravaud_2009a} and nuclear magnetic resonance (NMR) systems \citep{Moresi_2003,Raich_2004}. The magnetic field generated by a permanent magnet or a system of these is often calculated using a finite element method (FEM) approach, which is limited in precision and speed by the fineness of the finite element mesh. However, the magnetic field can be calculated through the demagnetization tensor, $\bfN$, which expresses the relation between the magnetic field and the magnetization, if the magnet is uniformly magnetized, as
\begin{equation}
\bfH=-\bfN\cdot\bfM,
\label{eq:demag}
\end{equation}
where $\bfH$ is the magnetic field at a given point and $\bfM$ is the magnetization. The demagnetization tensor $\bfN$ is a rank-2 symmetric tensor \citep{Moskowitz_1966}, which is known analytically for a range of geometries, e.g.\ for an ellipsoid \citep{Osborn_1945,Tejedor_1995}, a rectangular prism \citep{Joseph_1965,Smith_2010}, a finite cylinder \citep{Kraus_1973,Caciagli_2018}, a cylindrical tile \citep{Nielsen_2020,Slanovc_2022}, a tetrahedron \citep{Nielsen_2019}, cylindrical shells \citep{Beleggia_2009}, an equilateral triangular prism \citep{Tang_2005} and a hollow sphere \citep{PratCamps_2016}. The demagnetization tensor can also be expressed in Fourier space \citep{Beleggia_2003}. The magnetic field is found both inside and outside of the magnetized volume.

In this work we consider the magnetic field generated by what we term a circular piece. The circular piece is the part of a finite solid cylinder, with its axis orientated along the $z$-axis, that is cut off by two planes parallel to the $xz$- and $yz$-planes, respectively, so that its cross section is bounded by a circular arc, a vertical chord and a horizontal chord, as shown in Fig.~\ref{fig:geometry}. Together with its complement, i.e.\ the part of the bounding rectangle that lies outside the arc, the circular piece allows a cylindrical magnet to be split along arbitrary $x$- and $y$-values. This is needed e.g.\ when a cylindrical magnet is combined with rectangular parts in the same assembly, or when the field of a cylinder is required in a Cartesian subdivision, as is the case in micromagnetic and magnetostatic frameworks built on the analytical demagnetization tensor \citep{Bjork_2021}.

We derive the analytical expressions for the demagnetization tensor for a circular piece assuming the piece is homogeneously magnetized. The curved surface of the circular piece is identical to the outer surface of the cylindrical tile considered in Refs.~\citep{Nielsen_2020,Slanovc_2022}, while the rectangular surfaces parallel to the coordinate planes are identical to the surfaces of the rectangular prism \citep{Smith_2010}. However, the top and bottom surfaces of the circular piece, bounded by two edges parallel to the coordinate axes and a circular edge connecting these, are not found in other geometries. The computer code of this model, written in Fortran with Matlab and Python interfaces, is publicly available as a part of the MagTense framework \citep{MagTense}.

\section{The magnetic field of a circular piece}
\label{sec:model}
\begin{figure}[!t]
\centering
\begin{subfigure}[b]{0.53\textwidth}
\centering
\begin{tikzpicture}[scale=5]
\draw[thick] (9.66mm,2.59mm) arc (15:70:10mm);
\draw[thick] (9.66mm,2.59mm) -- (3.42mm,2.59mm);
\draw[thick] (3.42mm,2.59mm) -- (3.42mm,9.4mm);
\draw[dashed] (9.66mm,9.4mm) -- (3.42mm,9.4mm);
\draw[dashed] (9.66mm,9.4mm) -- (9.66mm,2.59mm);
\draw (9.66mm,2.59mm) -- (0,0);
\draw (3.42mm,9.4mm) -- (0,0);
\draw[dashed] (0,0) -- (7.37mm,6.76mm);
\draw[->] (0,0) -- (11.5mm,0mm) node[right]{$x$};
\draw[->] (0,0) -- (0mm,11.5mm) node[above]{$y$};
\draw (1.8mm,0) arc (0:15:1.8mm);
\draw (1.8mm,-0.5mm) node {$\theta_1$};
\draw (2.8mm,0) arc (0:42.5:2.8mm);
\draw (2.8mm,-0.5mm) node {$\theta_0$};
\draw (3.8mm,0) arc (0:70:3.8mm);
\draw (3.8mm,-0.5mm) node {$\theta_2$};
\draw (8.35mm,7.65mm) node {$\delta\theta$};
\draw (2.0mm,7mm) node {$R$};
\fill (9.66mm,2.59mm) circle (0.12mm) node[below right]{$1=(x_2,y_1)$};
\fill (3.42mm,9.4mm) circle (0.12mm) node[above,xshift=-2mm]{$2=(x_1,y_2)$};
\fill (3.42mm,2.59mm) circle (0.12mm) node[above right,xshift=4.5mm]{$3=(x_1,y_1)$};
\fill (9.66mm,9.4mm) circle (0.12mm) node[above right]{$3'=(x_2,y_2)$};
\end{tikzpicture}
\caption{}
\label{fig:geometry2d}
\end{subfigure}\hfill
\begin{subfigure}[b]{0.45\textwidth}
\centering
\begin{tikzpicture}[scale=4.6]
\draw[->] (0.00mm,0.00mm) -- (11.88mm,-0.84mm) node[right]{$x$};
\draw[->] (0.00mm,0.00mm) -- (3.34mm,11.88mm) node[right]{$y$};
\draw[->] (0.00mm,0.00mm) -- (0.00mm,6.50mm) node[above]{$z$};
\draw[dashed] (9.93mm,1.91mm) -- (9.87mm,2.11mm) -- (9.80mm,2.30mm) -- (9.71mm,2.50mm) -- (9.60mm,2.69mm) -- (9.49mm,2.88mm) -- (9.35mm,3.07mm) -- (9.20mm,3.26mm) -- (9.04mm,3.44mm) -- (8.86mm,3.62mm) -- (8.67mm,3.79mm) -- (8.46mm,3.96mm) -- (8.24mm,4.13mm) -- (8.01mm,4.29mm) -- (7.76mm,4.45mm) -- (7.50mm,4.60mm) -- (7.23mm,4.75mm) -- (6.95mm,4.89mm) -- (6.66mm,5.03mm) -- (6.36mm,5.16mm) -- (6.04mm,5.28mm) -- (5.72mm,5.40mm) -- (5.39mm,5.51mm) -- (5.04mm,5.62mm) -- (4.69mm,5.71mm);
\draw[dashed] (4.69mm,5.71mm) -- (3.75mm,2.34mm);
\draw[dashed] (4.69mm,5.71mm) -- (4.69mm,9.18mm);
\fill[gray!15] (9.93mm,5.37mm) -- (9.87mm,5.57mm) -- (9.80mm,5.77mm) -- (9.71mm,5.96mm) -- (9.60mm,6.16mm) -- (9.49mm,6.35mm) -- (9.35mm,6.53mm) -- (9.20mm,6.72mm) -- (9.04mm,6.90mm) -- (8.86mm,7.08mm) -- (8.67mm,7.26mm) -- (8.46mm,7.43mm) -- (8.24mm,7.60mm) -- (8.01mm,7.76mm) -- (7.76mm,7.92mm) -- (7.50mm,8.07mm) -- (7.23mm,8.22mm) -- (6.95mm,8.36mm) -- (6.66mm,8.49mm) -- (6.36mm,8.62mm) -- (6.04mm,8.75mm) -- (5.72mm,8.87mm) -- (5.39mm,8.98mm) -- (5.04mm,9.08mm) -- (4.69mm,9.18mm) -- (3.75mm,5.81mm) -- (9.93mm,5.37mm) -- cycle;
\draw[thick] (3.75mm,2.34mm) -- (9.93mm,1.91mm);
\draw[thick] (9.93mm,1.91mm) -- (9.93mm,5.37mm);
\draw[thick] (3.75mm,2.34mm) -- (3.75mm,5.81mm);
\draw[thick] (9.93mm,5.37mm) -- (9.87mm,5.57mm) -- (9.80mm,5.77mm) -- (9.71mm,5.96mm) -- (9.60mm,6.16mm) -- (9.49mm,6.35mm) -- (9.35mm,6.53mm) -- (9.20mm,6.72mm) -- (9.04mm,6.90mm) -- (8.86mm,7.08mm) -- (8.67mm,7.26mm) -- (8.46mm,7.43mm) -- (8.24mm,7.60mm) -- (8.01mm,7.76mm) -- (7.76mm,7.92mm) -- (7.50mm,8.07mm) -- (7.23mm,8.22mm) -- (6.95mm,8.36mm) -- (6.66mm,8.49mm) -- (6.36mm,8.62mm) -- (6.04mm,8.75mm) -- (5.72mm,8.87mm) -- (5.39mm,8.98mm) -- (5.04mm,9.08mm) -- (4.69mm,9.18mm) -- (3.75mm,5.81mm) -- (9.93mm,5.37mm) -- cycle;
\draw[dotted] (0.00mm,1.30mm) -- (3.75mm,2.34mm);
\draw[dotted] (0.00mm,3.03mm) -- (3.75mm,4.07mm);
\draw[dotted] (0.00mm,4.76mm) -- (3.75mm,5.81mm);
\draw (0.00mm,1.30mm) node[left]{$z_1$};
\draw (0.00mm,3.03mm) node[left]{$z_0$};
\draw (0.00mm,4.76mm) node[left]{$z_2$};
\draw[<->] (11.04mm,1.58mm) -- (11.04mm,5.04mm) node[midway,right]{$\delta z$};
\fill (9.93mm,5.37mm) circle (0.12mm) node[above left]{$1$};
\fill (4.69mm,9.18mm) circle (0.12mm) node[above left]{$2$};
\fill (3.75mm,5.81mm) circle (0.12mm) node[above left]{$3$};
\end{tikzpicture}
\caption{}
\label{fig:geometry3d}
\end{subfigure}
\caption{(a) The 2D cross section perpendicular to the cylinder axis of the circular piece considered in this work, for a piece in the first quadrant, and (b) the piece in 3D. The piece (solid outline) is bounded by the arc of radius $R$ from point 1 at the angle $\theta_1$ to point 2 at the angle $\theta_2$, by the vertical chord $2\to3$ at $x'=x_1$ and by the horizontal chord $3\to1$ at $y'=y_1$. The complementary piece of Sec.~\ref{sec:inverted} is bounded by the same arc and by the dashed chords through the corner $3'=(x_2,y_2)$. Both pieces extend from $z_1$ to $z_2$ in the $z$-direction, as shown in (b), where the hidden edges are dashed and the dotted lines mark the heights $z_1$, $z_0$ and $z_2$.}
\label{fig:geometry}
\end{figure}
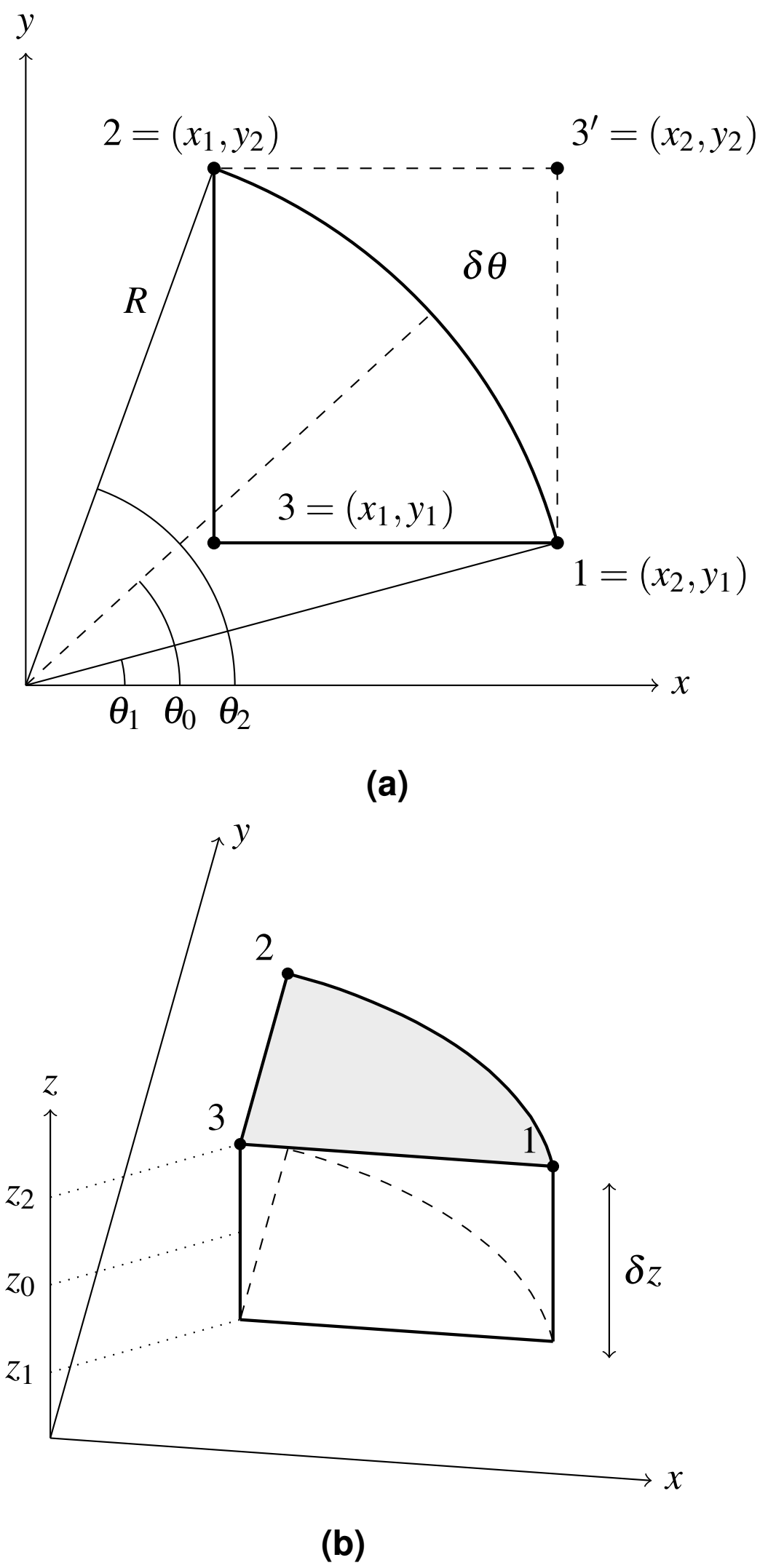

We consider a circular piece in 3D composed of five surfaces. As written above, the cylinder axis coincides with the $z$-axis and the cut planes are parallel to the $xz$- and $yz$-planes. The circular piece is completely specified by five parameters, namely the radius $R$ of the cylinder, the angle $\theta_0$ from the $x$-axis to the middle of the arc, the angular span of the arc $\delta\theta$, and the position $z_0$ and height $\delta z$ of the piece along the $z$-axis, as illustrated in Fig.~\ref{fig:geometry}. We define
\begin{eqnarray}
\theta_1&=&\theta_0-\delta\theta/2,\qquad \theta_2=\theta_0+\delta\theta/2,\qquad \nonumber\\ z_1&=&z_0-\delta z/2,\qquad z_2=z_0+\delta z/2 ,
\end{eqnarray}
and the coordinates of the rectangle bounding the cross section in the $xy$-plane are
\begin{eqnarray}
x_1&=&R\cos\theta_2,\qquad x_2=R\cos\theta_1,\nonumber\\ y_1&=&R\sin\theta_1,\qquad y_2=R\sin\theta_2 ,
\label{eq:corners}
\end{eqnarray}
so that the three corner points of the cross section are $1=(x_2,y_1)$ at the angle $\theta_1$, $2=(x_1,y_2)$ at the angle $\theta_2$ and $3=(x_1,y_1)$, see Fig.~\ref{fig:geometry}. The cross section is bounded by the arc of the circle from point 1 to point 2, the vertical chord from 2 to 3 at $x'=x_1$ and the horizontal chord from 3 to 1 at $y'=y_1$. In the $z$-direction the piece extends from $z_1$ to $z_2$. We assume the piece to be homogeneously magnetized with the magnetization $\bfM=(M_x,M_y,M_z)$. Our objective is to find the magnetic field, $\bfH$, at any point $\bfr=(x,y,z)$, be it inside or outside the circular piece. 

In the following the derivation is carried out for a piece whose arc lies in the first quadrant, i.e.\ $0<\theta_1<\theta_2<\pi/2$, so that $x_1<x_2$ and $y_1<y_2$. If an arc crosses a coordinate axis, the piece must be split at the axis into two pieces whose fields are then added. Pieces in the other three quadrants can be obtained by reflection, as described in Sec.~\ref{sec:quadrants}.

We are only interested in the magnetostatic limit, and assume no free currents. The magnetic field is then irrotational and can be written as the gradient of the magnetic scalar potential, $\bfH=-\nabla\phi_M$. For a homogeneously magnetized body the magnetic charge volume density $\rho_M=-\nabla\cdot\bfM$ vanishes, and the scalar potential is produced only by the magnetic charge surface density $\sigma_M=\nhat\cdot\bfM$ on the enclosing surface $S'$, where $\nhat$ is the outward unit normal of the surface. Following Jackson \citep{Jackson_1999} the potential is
\begin{equation}
\phi_M(\bfr)=\frac{1}{4\pi}\oint_{S'}\frac{\nhat(\bfr')\cdot\bfM}{\|\bfr-\bfr'\|}\,\ud S' .
\label{eq:phi}
\end{equation}
Note that there are two sets of coordinates. The coordinates marked with a $'$ are the coordinates of the surface that creates the magnetic field, whereas the non-marked coordinates are those of the point at which the field is evaluated. Note that the magnetic scalar potential has been found analytically for the prism \citep{James_2026} and the cylinder tile \citep{Bjork_2025}. The magnetic field is obtained as the negative gradient of Eq.~\eqref{eq:phi}. As the gradient operator and the integration are with respect to different sets of coordinates, the order of these is interchangeable, and as the magnetization is constant it can be moved outside the integral. We have
\begin{equation}
\bfH(\bfr)=-\bfN(\bfr)\cdot\bfM,\qquad
N_{ij}(\bfr)=\frac{1}{4\pi}\oint_{S'} n'_j\,\partial_i D\,\ud S',
\label{eq:Ndef}
\end{equation}
with
\begin{equation}
D(x',y',z')=\frac{1}{\|\bfr-\bfr'\|}=\frac{1}{\sqrt{(x-x')^2+(y-y')^2+(z-z')^2}} ,
\label{eq:Ddef}
\end{equation}
where $\partial_i$ denotes the partial derivative with respect to the $i$-th coordinate of the field point, $i=x,y,z$, and $n'_j$ is the $j$-th component of $\nhat(\bfr')$. In the following the arguments of $D$ are only written out when the dependence on the source point needs to be made explicit. We remark that Eq.~\eqref{eq:Ndef} gives the field both inside and outside the piece. The tensor is symmetric and its trace is $1$ for field points inside the body and $0$ outside \citep{Moskowitz_1966}.

The components of the tensor are computed by integrating along the entire closed surface of the circular piece. There are four unique surfaces: the arc surface at $r'=R$, the vertical surface at $x'=x_1$, the horizontal surface at $y'=y_1$ and the two end surfaces at $z'=z_1$ and $z'=z_2$. Only the column of $\bfN$ that multiplies the magnetization component along the surface normal is non-zero for a given surface. The total tensor is the sum of the contributions from the five surfaces,
\begin{equation}
\bfN=\bfN^{(1)}+\bfN^{(2)}+\bfN^{(3)}+\bfN^{(4,5)}.
\label{eq:Nsum}
\end{equation}
Thus, as each surface only couples to the magnetization component along its normal, the first column of the full tensor, $N_{ix}$, is the sum of the arc and the vertical surface, the second column, $N_{iy}$, is the sum of the arc and the horizontal surface, and the third column, $N_{iz}$, is given by the end surfaces alone. We note that the symmetry $N_{ij}=N_{ji}$ is not manifest in this decomposition: e.g.\ $N_{xy}$ is the sum of an arc term and a horizontal-surface term, while $N_{yx}$ is the sum of a different arc term and a vertical-surface term, and only the sums agree.

\subsection{Surface 1 -- the arc surface}
On the arc surface $\bfr'=(R\cos\theta',R\sin\theta',z')$, the outward normal is $\nhat'=\rhat=(\cos\theta',\sin\theta',0)$ and the surface element is $\ud S'=R\,\ud\theta'\,\ud z'$, with $\theta_1\le\theta'\le\theta_2$ and $z_1\le z'\le z_2$. The magnetic charge surface density is $\sigma_M=M_x\cos\theta'+M_y\sin\theta'$, and the elements of the tensor are
\begin{eqnarray}
N^{(1)}_{ix}=\frac{R}{4\pi}\int_{z_1}^{z_2}\!\!\int_{\theta_1}^{\theta_2}\cos\theta'\,\partial_i D\,\ud\theta'\,\ud z',\nonumber\\
N^{(1)}_{iy}=\frac{R}{4\pi}\int_{z_1}^{z_2}\!\!\int_{\theta_1}^{\theta_2}\sin\theta'\,\partial_i D\,\ud\theta'\,\ud z',
\label{eq:arcint}
\end{eqnarray}
for $i=x,y,z$, while $N^{(1)}_{iz}=0$. These six integrals are identical to the integrals over the outer arc surface of a cylindrical tile, see Refs.~\citep{Nielsen_2020,Slanovc_2022}.

\subsection{Surface 2 -- the vertical surface}
The vertical surface is defined for constant $x'=x_1$ and lies in the $y'z'$-plane, with $y_1\le y'\le y_2$ and $z_1\le z'\le z_2$. The outward normal is $\nhat'=-\xhat$, so that $\sigma_M=-M_x$ and
\begin{equation}
N^{(2)}_{ix}=-\frac{1}{4\pi}\int_{z_1}^{z_2}\!\!\int_{y_1}^{y_2}\partial_i D\,\ud y'\,\ud z'\Big|_{x'=x_1},
\label{eq:vertint}
\end{equation}
while $N^{(2)}_{iy}=N^{(2)}_{iz}=0$. This integral is identical to that over a surface of the rectangular prism, see Refs. \citep{Joseph_1965,Smith_2010}.

\subsection{Surface 3 -- the horizontal surface}
The horizontal surface is defined for constant $y'=y_1$ and lies in the $x'z'$-plane, with $x_1\le x'\le x_2$ and $z_1\le z'\le z_2$. The outward normal is $\nhat'=-\yhat$, so that $\sigma_M=-M_y$ and
\begin{equation}
N^{(3)}_{iy}=-\frac{1}{4\pi}\int_{z_1}^{z_2}\!\!\int_{x_1}^{x_2}\partial_i D\,\ud x'\,\ud z'\Big|_{y'=y_1},
\label{eq:horint}
\end{equation}
while $N^{(3)}_{ix}=N^{(3)}_{iz}=0$. This integral is also identical to that over a surface of the rectangular prism \citep{Joseph_1965,Smith_2010}.

\subsection{Surfaces 4 and 5 -- the end surfaces}
The two end surfaces are defined for constant $z'=z_2$, with the normal $\nhat'=\zhat$, and $z'=z_1$ with the normal $\nhat'=-\zhat$. Both are the cross section $A$ shown in Fig.~\ref{fig:geometry}. This can be described either as $y_1\le y'\le y_2$, $x_1\le x'\le\sqrt{R^2-y'^2}$, i.e.\ from the vertical chord to the arc, or as $x_1\le x'\le x_2$, $y_1\le y'\le\sqrt{R^2-x'^2}$, i.e.\ from the horizontal chord to the arc. Only $M_z$ contributes, and
\begin{equation}
N^{(4,5)}_{iz}=\frac{1}{4\pi}\left(\iint_A\partial_i D\,\ud x'\,\ud y'\Big|_{z'=z_2}-\iint_A\partial_i D\,\ud x'\,\ud y'\Big|_{z'=z_1}\right),
\label{eq:endint}
\end{equation}
while $N^{(4,5)}_{ix}=N^{(4,5)}_{iy}=0$.

\section{Evaluating the integrals}
\label{sec:evaluating}
The integrals over the arc surface and over the two planar chord surfaces have been evaluated previously for the cylindrical tile \citep{Nielsen_2020,Slanovc_2022} and the rectangular prism \citep{Joseph_1965,Smith_2010}, respectively, and we state these results in Secs.~\ref{sec:arc} and \ref{sec:chords} in the notation defined in this work. The integrals over the end surfaces are evaluated in Sec.~\ref{sec:caps}, where it is shown that they can be expressed through the same functions. Throughout we use the double difference notation
\begin{equation}
\begin{aligned}
\bx{f(p,q)}{p_1}{q_1}{p_2}{q_2}=\;&f(p_2,q_2)-f(p_1,q_2)\\
&-f(p_2,q_1)+f(p_1,q_1)
\end{aligned}
\label{eq:boxdef}
\end{equation}
for the evaluation of a double integral.

\subsection{The arc surface}
\label{sec:arc}
The two integrals of Eq.~\eqref{eq:arcint}, each for $i=x,y,z$, i.e.\ six integrals in total, were evaluated in Refs.~\citep{Nielsen_2020,Slanovc_2022} by means of the rotation and translation trick of Ref.~\citep{Varga_1998}, which expresses the integrals as functions of the distance of the field point from the $z$-axis and the primed coordinates only. Along the $z$-axis this is achieved by a translation, i.e.\ by measuring the $z$-coordinate of the source point relative to that of the field point,
\begin{equation}
\Dz=z'-z,\qquad \Dz_1=z_1-z,\qquad \Dz_2=z_2-z ,
\label{eq:Dzdef}
\end{equation}
where $\Dz_1$ and $\Dz_2$ are the positions of the two end surfaces relative to the field point. These depend on the field point, whereas their difference $\Dz_2-\Dz_1=\delta z$ is the height of the piece. In the $xy$-plane the field point is rotated about the $z$-axis so that it lies on the $x$-axis. With the azimuth and the distance from the axis of the field point,
\begin{equation}
\varphi=\atantwo(y,x)\in[0,2\pi),\qquad \xr=\sqrt{x^2+y^2},
\end{equation}
the rotation by $-\varphi$ maps the field point to $(\xr,0,z)$ and the angle $\theta'$ on the arc to the rotated angle $\psi=\theta'-\varphi$, so that the integration limits become
\begin{equation}
\psi_1=\theta_1-\varphi,\qquad \psi_2=\theta_2-\varphi ,
\label{eq:thetalimits}
\end{equation}
and the magnetic field is finally rotated back about the $z$-axis with the angle $\varphi$. In the rotated frame the distance between the field point and a point on the arc surface is $1/D=\hA(R,\xr,\psi,\Dz)$, where, following Ref.~\citep{Nielsen_2020}, we introduce the helper functions
\begin{equation}
\begin{aligned}
\hA(R,\xr,\psi,\Dz)&=\sqrt{R^2-2\xr R\cos \psi+\xr^2+\Dz^2}, \\
\hC(R,\xr)&=\frac{4R\xr}{(R+\xr)^2},\\
\hD(\psi)&=\cos\frac{\psi}{2}, \\
\hE(R,\xr,\Dz)&=2\sqrt{\frac{R\xr}{(R+\xr)^2+\Dz^2}} .
\end{aligned}
\label{eq:helpers}
\end{equation}
The six integrals of Eq.~\eqref{eq:arcint} then reduce to integrals over the six integrands
\begin{equation}
\begin{aligned}
g_\mathcal{A}&=\frac{\cos \psi\,(\xr-R\cos \psi)}{\hA^3}, &
g_\mathcal{B}&=\frac{\sin \psi\,(\xr-R\cos \psi)}{\hA^3}, \\
g_\mathcal{C}&=\frac{\cos \psi\,R\sin \psi}{\hA^3}, &
g_\mathcal{D}&=\frac{\sin \psi\,R\sin \psi}{\hA^3}, \\
g_\mathcal{E}&=\frac{\cos \psi\,\Dz}{\hA^3}, &
g_\mathcal{F}&=\frac{\sin \psi\,\Dz}{\hA^3},
\end{aligned}
\label{eq:gdef}
\end{equation}
over $\psi_1\le \psi\le \psi_2$ and $\Dz_1\le\Dz\le\Dz_2$. Denoting by $\mathcal{A},\dots,\mathcal{F}$ the integrals of these integrands, i.e.\ the functions with \\ $\partial^2\mathcal{A}/\partial\psi\partial\Dz=g_\mathcal{A}$ etc., the double integrals are the double differences of Eq.~\eqref{eq:boxdef} over the rectangle $[\psi_1,\psi_2]\times[\Dz_1,\Dz_2]$,
\begin{equation}
\bxs{\mathcal{A}}=\int_{\Dz_1}^{\Dz_2}\!\!\int_{\psi_1}^{\psi_2}g_\mathcal{A}\,\ud \psi\,\ud\Dz=\bx{\mathcal{A}(\psi,\Dz)}{\psi_1}{\Dz_1}{\psi_2}{\Dz_2},
\label{eq:boxA}
\end{equation}
and likewise for $\bxs{\mathcal{B}},\dots,\bxs{\mathcal{F}}$. Rotating the field back and collecting the coefficients of $M_x$ and $M_y$, the contribution of the arc surface to the tensor becomes
\begin{align}
N^{(1)}_{xx}&=-\frac{R}{4\pi}\biggl(\cos^2\varphi\,\bxs{\mathcal{A}}-\sin\varphi\cos\varphi\,\bxs{\mathcal{B}}\nonumber\\
&\qquad+\sin\varphi\cos\varphi\,\bxs{\mathcal{C}}-\sin^2\varphi\,\bxs{\mathcal{D}}\biggr),\nonumber\\
N^{(1)}_{xy}&=-\frac{R}{4\pi}\biggl(\sin\varphi\cos\varphi\,\bxs{\mathcal{A}}+\cos^2\varphi\,\bxs{\mathcal{B}}\nonumber\\
&\qquad+\sin^2\varphi\,\bxs{\mathcal{C}}+\sin\varphi\cos\varphi\,\bxs{\mathcal{D}}\biggr),\nonumber\\
N^{(1)}_{yx}&=-\frac{R}{4\pi}\biggl(\sin\varphi\cos\varphi\,\bxs{\mathcal{A}}-\sin^2\varphi\,\bxs{\mathcal{B}}\nonumber\\
&\qquad-\cos^2\varphi\,\bxs{\mathcal{C}}+\sin\varphi\cos\varphi\,\bxs{\mathcal{D}}\biggr),\nonumber\\
N^{(1)}_{yy}&=-\frac{R}{4\pi}\biggl(\sin^2\varphi\,\bxs{\mathcal{A}}+\sin\varphi\cos\varphi\,\bxs{\mathcal{B}}\nonumber\\
&\qquad-\sin\varphi\cos\varphi\,\bxs{\mathcal{C}}-\cos^2\varphi\,\bxs{\mathcal{D}}\biggr),\nonumber\\
N^{(1)}_{zx}&=\frac{R}{4\pi}\biggl(\cos\varphi\,\bxs{\mathcal{E}}-\sin\varphi\,\bxs{\mathcal{F}}\biggr),\nonumber\\
N^{(1)}_{zy}&=\frac{R}{4\pi}\biggl(\sin\varphi\,\bxs{\mathcal{E}}+\cos\varphi\,\bxs{\mathcal{F}}\biggr).
\label{eq:NIII}
\end{align}

The integrals are given in Ref.~\citep{Nielsen_2020,Slanovc_2022} in terms of the helper functions of Eq.~\eqref{eq:helpers} and the incomplete elliptic integrals of the first, second and third kind, denoted $F$, $E$ and $\Pi$, respectively, which are defined as
\begin{equation}
\begin{aligned}
F(s,k)&=\int_0^{s}\frac{\ud t}{\sqrt{(1-t^2)(1-k^2t^2)}},\\
E(s,k)&=\int_0^{s}\frac{\sqrt{1-k^2t^2}}{\sqrt{1-t^2}}\,\ud t,\\
\Pi(s,\nu,k)&=\int_0^{s}\frac{\ud t}{(1-\nu t^2)\sqrt{(1-t^2)(1-k^2t^2)}} .
\end{aligned}
\label{eq:ellip}
\end{equation}
In this convention, which is that of Maple, the first argument is the sine of the amplitude, $k$ is the modulus, i.e.\ the parameter of Legendre's form is $m=k^2$, and the characteristic $\nu$ enters as $1-\nu\sin^2$. All three functions are odd in $s$. In the integrals the arguments are $s=\hD(\psi)$, $\nu=\hC(R,\xr)$ and $k=\hE(R,\xr,\Dz)$, so that $\nu\le1$ with equality for $\xr=R$, and $k<1$ for $\Dz\ne0$ or $\xr\ne R$. The integrals are, for $0<\psi<2\pi$,
\begin{align}
\mathcal{A}(\psi,\Dz)&=\frac{\Dz}{2R\xr^2(R+\xr)\sqrt{(R+\xr)^2+\Dz^2}}\nonumber\\
&\quad\times\Bigl(-(\xr-R)(R^2+\xr^2)\,\Pi(\hD,\hC,\hE)\nonumber\\
&\qquad+(R+\xr)\bigl((R+\xr)^2+\Dz^2\bigr)E(\hD,\hE)\nonumber\\
&\qquad-(R+\xr)(2R^2+\Dz^2)F(\hD,\hE)\Bigr),\label{eq:GA}\\
\mathcal{B}(\psi,\Dz)&=\frac{1}{4R\xr^2}\Bigl((R^2-\xr^2)\ln\frac{\hA+\Dz}{\hA-\Dz}+2\Dz\hA\Bigr),\label{eq:GB}\\
\mathcal{C}(\psi,\Dz)&=-\frac{1}{2R\xr^2}\Bigl((R^2+\xr^2)\ln(\hA+\Dz)+\Dz\hA\Bigr),\label{eq:GC}\\
\mathcal{D}(\psi,\Dz)&=\frac{\Dz}{2R\xr^2\sqrt{(R+\xr)^2+\Dz^2}}\Bigl((\xr-R)^2\,\Pi(\hD,\hC,\hE)\nonumber\\
&\qquad+\bigl((R+\xr)^2+\Dz^2\bigr)E(\hD,\hE)\nonumber\\
&\qquad-(2\xr^2+2R^2+\Dz^2)F(\hD,\hE)\Bigr),\label{eq:GD}\\
\mathcal{E}(\psi,\Dz)&=-\frac{1}{R\xr\sqrt{(R+\xr)^2+\Dz^2}}\nonumber\\
&\quad\times\Bigl(\bigl((R+\xr)^2+\Dz^2\bigr)E(\hD,\hE)\nonumber\\
&\qquad-(R^2+\xr^2+\Dz^2)F(\hD,\hE)\Bigr),\label{eq:GE}\\
\mathcal{F}(\psi,\Dz)&=-\frac{\hA}{R\xr}.\label{eq:GF}
\end{align}
In the notation of Ref.~\citep{Nielsen_2020}, $\mathcal{A}$, $\mathcal{B}$, $\mathcal{C}$, $\mathcal{D}$, $\mathcal{E}$ and $\mathcal{F}$ are $-\mathcal{I}$, $-\mathcal{G}$, $\mathcal{D}/r_s$, $\mathcal{A}$, $\mathcal{E}$ and $\mathcal{J}$, respectively, where the signs follow from $\partial_x D=-(x-x')D^3$ and where we note that the expression given for $\mathcal{D}$ in Ref.~\citep{Nielsen_2020} corresponds to $r_s$ times the integral defined there. For $\pi<\psi<2\pi$ the first argument $\hD$ is negative and the elliptic integrals are evaluated as odd functions of it; $\mathcal{A}$, $\mathcal{D}$ and $\mathcal{E}$ vanish at $\psi=\pi$ and are continuous there, so the double differences may be used directly whenever $0\le \psi_1<\psi_2\le2\pi$.

\subsubsection{Extension to negative angles}
\label{sec:oddext}
The limits $\psi_1$ and $\psi_2$ of Eq.~\eqref{eq:thetalimits} lie in $(-2\pi,2\pi)$ and may have either sign, and they straddle $\psi=0$ precisely when the azimuth of the field point lies within the angular span of the piece, $\theta_1<\varphi<\theta_2$. The integrands $g_\mathcal{B}$, $g_\mathcal{C}$ and $g_\mathcal{F}$ are odd in $\psi$, and $\mathcal{B}$, $\mathcal{C}$ and $\mathcal{F}$, which depend on $\psi$ only through $\cos \psi$, are even and valid integrals for all $\psi$. The integrands $g_\mathcal{A}$, $g_\mathcal{D}$ and $g_\mathcal{E}$ are even in $\psi$, so their integrals must be odd. However, $\mathcal{A}$, $\mathcal{D}$ and $\mathcal{E}$ are even functions of $\psi$ that do not vanish at $\psi=0$, where $\hD=1$ and they are built from the complete elliptic integrals. The odd extension $\sgn(\sin \psi/2)\mathcal{A}$ used in Ref.~\citep{Nielsen_2020} therefore has a jump of $2\mathcal{A}(0,\Dz)$ at $\psi=0$ and cannot be evaluated across it. The continuous odd integral is obtained by subtracting the value at the origin,
\begin{equation}
\begin{aligned}
\tilde{\mathcal{A}}(\psi,\Dz)&=\sgn(\psi)\Bigl(\mathcal{A}(|\psi|,\Dz)-\mathcal{A}(0,\Dz)\Bigr),\\
&\qquad -2\pi<\psi<2\pi ,
\end{aligned}
\label{eq:oddext}
\end{equation}
and likewise for $\tilde{\mathcal{D}}$ and $\tilde{\mathcal{E}}$, while $\tilde{\mathcal{B}}=\mathcal{B}$, $\tilde{\mathcal{C}}=\mathcal{C}$ and $\tilde{\mathcal{F}}=\mathcal{F}$. We note that $\mathcal{A}(0,\Dz)$ depends on $\Dz$, so the subtraction does not cancel in the double difference. The double differences in Eq.~\eqref{eq:NIII} are then to be evaluated with the extended integrals, i.e.\ Eq.~\eqref{eq:boxA} is replaced by
\begin{equation}
\bxs{\mathcal{A}}=\bx{\tilde{\mathcal{A}}(\psi,\Dz)}{\psi_1}{\Dz_1}{\psi_2}{\Dz_2},
\label{eq:Ibox}
\end{equation}
and likewise for the remaining five double differences. Equation \eqref{eq:oddext} is equivalent to splitting the integration over $\ud \psi$ at $\psi=0$ and using the symmetry of the integrand on the negative part.

\subsection{The vertical and horizontal surfaces}
\label{sec:chords}
The vertical and horizontal surfaces are rectangles parallel to the coordinate planes, and the integrals over them are those over the faces of the rectangular prism \citep{Joseph_1965,Smith_2010}. In the present notation the integrals for the vertical surface at $x'=x_1$ are
\begin{equation}
\begin{aligned}
\mathcal{G}(y',z')&=\arctan\frac{(y-y')(z-z')\,D(x_1,y',z')}{x-x_1},\\
\mathcal{H}(y',z')&=\ln\bigl(z'-z+1/D(x_1,y',z')\bigr),\\
\mathcal{I}(y',z')&=\ln\bigl(y'-y+1/D(x_1,y',z')\bigr),
\end{aligned}
\label{eq:chordvert}
\end{equation}
and for the horizontal surface at $y'=y_1$
\begin{equation}
\begin{aligned}
\mathcal{J}(x',z')&=\ln\bigl(z'-z+1/D(x',y_1,z')\bigr),\\
\mathcal{K}(x',z')&=\arctan\frac{(x-x')(z-z')\,D(x',y_1,z')}{y-y_1},\\
\mathcal{L}(x',z')&=\ln\bigl(x'-x+1/D(x',y_1,z')\bigr),
\end{aligned}
\label{eq:chordhor}
\end{equation}
with which Eqs.~\eqref{eq:vertint} and \eqref{eq:horint} evaluate to
\begin{equation}
\begin{aligned}
N^{(2)}_{xx}&=\frac{1}{4\pi}\bx{\mathcal{G}}{y_1}{z_1}{y_2}{z_2}, &
N^{(3)}_{xy}&=\frac{1}{4\pi}\bx{\mathcal{J}}{x_1}{z_1}{x_2}{z_2},\\
N^{(2)}_{yx}&=\frac{1}{4\pi}\bx{\mathcal{H}}{y_1}{z_1}{y_2}{z_2}, &
N^{(3)}_{yy}&=\frac{1}{4\pi}\bx{\mathcal{K}}{x_1}{z_1}{x_2}{z_2},\\
N^{(2)}_{zx}&=\frac{1}{4\pi}\bx{\mathcal{I}}{y_1}{z_1}{y_2}{z_2}, &
N^{(3)}_{zy}&=\frac{1}{4\pi}\bx{\mathcal{L}}{x_1}{z_1}{x_2}{z_2}.
\end{aligned}
\label{eq:chords}
\end{equation}
The expressions for the horizontal surface follow from those for the vertical surface by interchanging the roles of $x$ and $y$. The arctangents in $\mathcal{G}$ and $\mathcal{K}$ are singular for field points in the plane of the respective surface, $x=x_1$ or $y=y_1$, where the arctangent tends to $\pm\pi/2$ depending on the side from which the plane is approached.

\subsection{The end surfaces}
\label{sec:caps}
The end surfaces are not present in any other geometry for which the demagnetization tensor is known. We consider one end surface at height $z'$; the two surfaces enter Eq.~\eqref{eq:endint} with opposite signs. Inserting $\partial_x D=-(x-x')D^3$, and correspondingly for $y$ and $z$, and using that $\partial D/\partial x'=-\partial_x D$, the $x$-component is integrated first over $x'$ from the vertical chord to the arc,
\begin{align}
\iint_A\partial_x D\,\ud x'\,\ud y'
&=-\int_{y_1}^{y_2}\!\!\int_{x_1}^{\sqrt{R^2-y'^2}}(x-x')D^3\,\ud x'\,\ud y'\nonumber\\
&=\int_{y_1}^{y_2}\Bigl(D(x_1,y',z')\nonumber\\
&\qquad-D\bigl(\sqrt{R^2-y'^2},y',z'\bigr)\Bigr)\ud y' .
\label{eq:capx}
\end{align}
The first term is elementary,
\begin{equation}
\int_{y_1}^{y_2}D(x_1,y',z')\,\ud y'=\Bigl[\ln\bigl(y'-y+1/D(x_1,y',z')\bigr)\Bigr]_{y'=y_1}^{y_2},
\label{eq:lnM}
\end{equation}
while in the second term the arc enters through $\sqrt{R^2-y'^2}$ inside $D$. The $y$-component is treated in the same way with the roles of $x'$ and $y'$ interchanged, i.e.\ integrating first over $y'$ from the horizontal chord to the arc,
\begin{equation}
\begin{aligned}
\iint_A\partial_y D\,\ud x'\,\ud y'
&=\int_{x_1}^{x_2}\Bigl(D(x',y_1,z')\\
&\qquad-D\bigl(x',\sqrt{R^2-x'^2},z'\bigr)\Bigr)\ud x' .
\end{aligned}
\label{eq:capy}
\end{equation}
The first term is again elementary,
\begin{equation}
\int_{x_1}^{x_2}D(x',y_1,z')\,\ud x'=\Bigl[\ln\bigl(x'-x+1/D(x',y_1,z')\bigr)\Bigr]_{x'=x_1}^{x_2}.
\label{eq:lnN}
\end{equation}
For the $z$-component the inner integral is $\int D^3\,\ud x'=(x'-x)D/\bigl((y-y')^2+(z-z')^2\bigr)$, so that
\begin{align}
&\iint_A\partial_z D\,\ud x'\,\ud y'\nonumber\\
&\quad=\int_{y_1}^{y_2}\frac{\bigl(x-\sqrt{R^2-y'^2}\bigr)(z-z')}{(y-y')^2+(z-z')^2}\,D\bigl(\sqrt{R^2-y'^2},y',z'\bigr)\,\ud y'\nonumber\\
&\qquad-\int_{y_1}^{y_2}\frac{(x-x_1)(z-z')}{(y-y')^2+(z-z')^2}\,D(x_1,y',z')\,\ud y' .
\label{eq:capz}
\end{align}
The second term is elementary by means of the identity
\begin{equation}
\frac{\partial}{\partial y'}\arctan\frac{(x-x')(y-y')\,D}{z-z'}=-\frac{(x-x')(z-z')\,D}{(y-y')^2+(z-z')^2} ,
\label{eq:atanid}
\end{equation}
which gives
\begin{equation}
\begin{aligned}
&-\int_{y_1}^{y_2}\frac{(x-x_1)(z-z')}{(y-y')^2+(z-z')^2}\,D(x_1,y',z')\,\ud y'\\
&\quad=\Bigl[\arctan\frac{(x-x_1)(y-y')\,D(x_1,y',z')}{z-z'}\Bigr]_{y'=y_1}^{y_2} .
\end{aligned}
\label{eq:atanO}
\end{equation}
Collecting the terms, we define for an end surface at height $z'$
\begin{align}
\mathcal{M}(z')&=\Bigl[\ln\bigl(y'-y+1/D(x_1,y',z')\bigr)\Bigr]_{y'=y_1}^{y_2}\nonumber\\
&\quad-\int_{y_1}^{y_2}D\bigl(\sqrt{R^2-y'^2},y',z'\bigr)\,\ud y',\label{eq:QM}\\
\mathcal{N}(z')&=\Bigl[\ln\bigl(x'-x+1/D(x',y_1,z')\bigr)\Bigr]_{x'=x_1}^{x_2}\nonumber\\
&\quad-\int_{x_1}^{x_2}D\bigl(x',\sqrt{R^2-x'^2},z'\bigr)\,\ud x',\label{eq:QN}\\
\mathcal{O}(z')&=\int_{y_1}^{y_2}\frac{\bigl(x-\sqrt{R^2-y'^2}\bigr)(z-z')}{(y-y')^2+(z-z')^2}\nonumber\\
&\qquad\times D\bigl(\sqrt{R^2-y'^2},y',z'\bigr)\,\ud y'\nonumber\\
&\quad+\Bigl[\arctan\frac{(x-x_1)(y-y')\,D(x_1,y',z')}{z-z'}\Bigr]_{y'=y_1}^{y_2},\label{eq:QO}
\end{align}
which are Eqs.~\eqref{eq:capx}, \eqref{eq:capy} and \eqref{eq:capz}, respectively, so that the third column of the tensor, Eq.~\eqref{eq:endint}, becomes
\begin{equation}
\begin{aligned}
N_{xz}&=\frac{\mathcal{M}(z_2)-\mathcal{M}(z_1)}{4\pi}, &
N_{yz}&=\frac{\mathcal{N}(z_2)-\mathcal{N}(z_1)}{4\pi},\\
N_{zz}&=\frac{\mathcal{O}(z_2)-\mathcal{O}(z_1)}{4\pi}.
\end{aligned}
\label{eq:Ncaps}
\end{equation}
The remaining one-dimensional integrals in $\mathcal{M}$, $\mathcal{N}$ and $\mathcal{O}$, in which the arc enters through $\sqrt{R^2-y'^2}$ or $\sqrt{R^2-x'^2}$ inside $D$, can be expressed in closed form in terms of the elliptic integrals already introduced, as shown in the following.

\subsubsection{The integrals $\mathcal{M}$ and $\mathcal{N}$}
\label{sec:MN}
Along the arc the substitution $y'=R\sin\theta'$, with $\sqrt{R^2-y'^2}=R\cos\theta'$ and $\ud y'=R\cos\theta'\,\ud\theta'$, turns the inverse distance to a point on the arc into that of the arc surface,
\begin{equation*}
D(R\cos\theta',R\sin\theta',z')=1/\hA(R,\xr,\psi,\Dz),
\end{equation*}
with the rotated angle $\psi=\theta'-\varphi$ and $\Dz=z'-z$ of Sec.~\ref{sec:arc}, so that with the limits $\psi_1$ and $\psi_2$ of Eq.~\eqref{eq:thetalimits}
\begin{align}
&\int_{y_1}^{y_2}D\bigl(\sqrt{R^2-y'^2},y',z'\bigr)\,\ud y'
=R\int_{\psi_1}^{\psi_2}\frac{\cos(\psi+\varphi)}{\hA}\,\ud \psi\nonumber\\
&\quad=R\cos\varphi\int_{\psi_1}^{\psi_2}\frac{\cos \psi}{\hA}\,\ud \psi-R\sin\varphi\int_{\psi_1}^{\psi_2}\frac{\sin \psi}{\hA}\,\ud \psi ,
\label{eq:Msub}
\end{align}
and likewise, with $x'=R\cos\theta'$,
\begin{equation}
\begin{aligned}
&\int_{x_1}^{x_2}D\bigl(x',\sqrt{R^2-x'^2},z'\bigr)\,\ud x'\\
&\quad=R\sin\varphi\int_{\psi_1}^{\psi_2}\frac{\cos \psi}{\hA}\,\ud \psi+R\cos\varphi\int_{\psi_1}^{\psi_2}\frac{\sin \psi}{\hA}\,\ud \psi .
\end{aligned}
\label{eq:Nsub}
\end{equation}
The two integrals are already at hand as differentiating Eqs.~\eqref{eq:GE} and \eqref{eq:GF} with respect to $\psi$ gives
\begin{equation}
\frac{\partial\mathcal{E}}{\partial \psi}=-\frac{\cos \psi}{\hA},\qquad
\frac{\partial\mathcal{F}}{\partial \psi}=-\frac{\sin \psi}{\hA},
\label{eq:dEdu}
\end{equation}
With the odd extension $\tilde{\mathcal{E}}$ of Sec.~\ref{sec:oddext} and the single difference $[f]_{\psi_1}^{\psi_2}=f(\psi_2)-f(\psi_1)$ at fixed $\Dz$,
\begin{align}
\mathcal{M}(z')&=R\Bigl[\cos\varphi\,\tilde{\mathcal{E}}(\psi,\Dz)-\sin\varphi\,\mathcal{F}(\psi,\Dz)\Bigr]_{\psi_1}^{\psi_2}\nonumber\\
&\quad+\Bigl[\ln\bigl(y'-y+1/D(x_1,y',z')\bigr)\Bigr]_{y'=y_1}^{y_2},\label{eq:Mclosed}\\
\mathcal{N}(z')&=R\Bigl[\sin\varphi\,\tilde{\mathcal{E}}(\psi,\Dz)+\cos\varphi\,\mathcal{F}(\psi,\Dz)\Bigr]_{\psi_1}^{\psi_2}\nonumber\\
&\quad+\Bigl[\ln\bigl(x'-x+1/D(x',y_1,z')\bigr)\Bigr]_{x'=x_1}^{x_2}.\label{eq:Nclosed}
\end{align}
Inserting these into Eq.~\eqref{eq:Ncaps}, the differences between the two end surfaces turn the single differences into the double differences of Eq.~\eqref{eq:boxdef}, and comparing with Eqs.~\eqref{eq:NIII} and \eqref{eq:chords} we find, with $\mathcal{E}$ and $\mathcal{F}$ of Eqs.~\eqref{eq:GE} and \eqref{eq:GF} and $\mathcal{I}$ and $\mathcal{L}$ of Eqs.~\eqref{eq:chordvert} and \eqref{eq:chordhor},
\begin{equation}
\begin{aligned}
N_{xz}&=\frac{R}{4\pi}\biggl(\cos\varphi\,\bxs{\mathcal{E}}-\sin\varphi\,\bxs{\mathcal{F}}\biggr)+\frac{1}{4\pi}\bx{\mathcal{I}}{y_1}{z_1}{y_2}{z_2}\\
&=N^{(1)}_{zx}+N^{(2)}_{zx}=N_{zx},\\
N_{yz}&=\frac{R}{4\pi}\biggl(\sin\varphi\,\bxs{\mathcal{E}}+\cos\varphi\,\bxs{\mathcal{F}}\biggr)+\frac{1}{4\pi}\bx{\mathcal{L}}{x_1}{z_1}{x_2}{z_2}\\
&=N^{(1)}_{zy}+N^{(3)}_{zy}=N_{zy}.
\end{aligned}
\label{eq:Nsym}
\end{equation}
The symmetry of the tensor \citep{Moskowitz_1966} is thus obtained explicitly: the arc term of an end surface is the $z$-derivative term of the arc surface, and the chord term of an end surface is the $z$-derivative term of the corresponding chord surface.

\subsubsection{The integral $\mathcal{O}$}
\label{sec:O}
For the $z$-component we note that $\mathcal{O}(z')=\Dz\iint_A D^3\,\ud A'$, i.e.\ $\mathcal{O}(z')=\sgn(\Dz)\,\Omega_A$, where $\Omega_A$ is the solid angle subtended by the end surface at the field point. The cross section $A$ is the circular sector $S=\{0\le r'\le R,\ \theta_1\le\theta'\le\theta_2\}$ minus the two triangles $T_1$ with the vertices $O$, $1$, $3$ and $T_2$ with the vertices $O$, $3$, $2$, see Fig.~\ref{fig:geometry}, and since solid angles are additive,
\begin{equation}
\Omega_A=\Omega_S-\Omega_{T_1}-\Omega_{T_2}.
\label{eq:Omsplit}
\end{equation}
The solid angle of a plane triangle with vertices at the positions $\mathbf{p}_1$, $\mathbf{p}_2$ and $\mathbf{p}_3$ relative to the field point is given by Ref. \citep{VanOosterom_1983},
\begin{equation}
\begin{aligned}
&\Omega_T(\mathbf{p}_1,\mathbf{p}_2,\mathbf{p}_3)=2\,\atantwo\Bigl(\bigl|\mathbf{p}_1\cdot(\mathbf{p}_2\times\mathbf{p}_3)\bigr|,\\
&\qquad p_1p_2p_3+(\mathbf{p}_1\cdot\mathbf{p}_2)p_3+(\mathbf{p}_1\cdot\mathbf{p}_3)p_2+(\mathbf{p}_2\cdot\mathbf{p}_3)p_1\Bigr),
\end{aligned}
\label{eq:OmT}
\end{equation}
with $p_i=|\mathbf{p}_i|$. With the positions of the corner points of Fig.~\ref{fig:geometry} and of the origin relative to the field point,
\begin{equation}
\begin{aligned}
\mathbf{p}_O&=(-x,\,-y,\,\Dz), & \mathbf{p}_1&=(x_2-x,\,y_1-y,\,\Dz),\\
\mathbf{p}_2&=(x_1-x,\,y_2-y,\,\Dz), & \mathbf{p}_3&=(x_1-x,\,y_1-y,\,\Dz),
\end{aligned}
\end{equation}
the two triangles of Eq.~\eqref{eq:Omsplit} subtend
\begin{equation}
\Omega_{T_1}=\Omega_T(\mathbf{p}_O,\mathbf{p}_1,\mathbf{p}_3),\qquad
\Omega_{T_2}=\Omega_T(\mathbf{p}_O,\mathbf{p}_3,\mathbf{p}_2).
\label{eq:OmT12}
\end{equation}
For the sector we integrate first over $r'$, using
\begin{equation}
\begin{aligned}
&\int\frac{r'}{(r'^2-2\xr r'\cos \psi+\rho^2)^{3/2}}\,\ud r'\\
&\quad=\frac{\xr r'\cos \psi-\rho^2}{(\xr^2\sin^2\psi+\Dz^2)\sqrt{r'^2-2\xr r'\cos \psi+\rho^2}} ,
\end{aligned}
\label{eq:rint}
\end{equation}
where $\rho=\sqrt{\xr^2+\Dz^2}$ is the distance from the field point to the centre of the end surface, and then over $\psi=\theta'-\varphi$,
\begin{equation}
\begin{aligned}
\Omega_S&=|\Dz|\int_{\psi_1}^{\psi_2}\biggl(\frac{\xr R\cos \psi-\rho^2}{\hA\,(\xr^2\sin^2\psi+\Dz^2)}\\
&\qquad+\frac{\rho}{\xr^2\sin^2\psi+\Dz^2}\biggr)\ud \psi .
\end{aligned}
\label{eq:OmSint}
\end{equation}
The second term of Eq.~\eqref{eq:OmSint} is elementary. Its integral $\arctan(\rho\tan \psi/|\Dz|)$ has to be continued continuously across $\psi=\pm\pi/2$ and $\pm3\pi/2$, which we write, following Ref.~\citep{Slanovc_2022}, as
\begin{equation}
\begin{aligned}
\mathrm{Sc}(\psi,c)&=m\pi+\arctan\bigl(c\tan(\psi-m\pi)\bigr),\\
m&=\bigl\lfloor \psi/\pi+\tfrac12\bigr\rfloor .
\end{aligned}
\label{eq:Sc}
\end{equation}
For the first term we substitute $t=\hD(\psi)=\cos(\psi/2)$. With $\cos\psi=2t^2-1$ the helper functions of Eq.~\eqref{eq:helpers} give
\begin{equation}
\begin{aligned}
\ud \psi&=-\frac{2\,\ud t}{\sqrt{1-t^2}},\\
\hA&=\sqrt{(R+\xr)^2+\Dz^2}\,\sqrt{1-\hE^2t^2},\\
\xr^2\sin^2\psi+\Dz^2&=4\xr^2t^2(1-t^2)+\Dz^2\\
&=-4\xr^2(t^2-\tau_+)(t^2-\tau_-),\\
\tau_\pm&=\frac{\xr\pm\rho}{2\xr} ,
\end{aligned}
\end{equation}
so that a partial fraction decomposition in $t^2$ and the definition of $\Pi$, Eq. \eqref{eq:ellip}, give a combination of two elliptic integrals of the third kind with the characteristics $\nu_\pm=1/\tau_\pm$. The same characteristics appear in the end-surface terms of the cylindrical tile in Ref.~\citep{Slanovc_2022}. The result is
\begin{equation}
\begin{aligned}
\Omega_S&=\Bigl[\mathrm{Sc}\bigl(\psi,\rho/|\Dz|\bigr)-\frac{|\Dz|}{2\xr\rho\sqrt{(R+\xr)^2+\Dz^2}}\\
&\qquad\times\Bigl((\alpha+\beta\nu_+)\tilde{\Pi}(\hD,\nu_+,\hE)\\
&\qquad\qquad-(\alpha+\beta\nu_-)\tilde{\Pi}(\hD,\nu_-,\hE)\Bigr)\Bigr]_{\psi_1}^{\psi_2},
\end{aligned}
\label{eq:OmS}
\end{equation}
with
\begin{equation}
\nu_\pm=\frac{2\xr}{\xr\pm\rho},\qquad \alpha=2\xr R,\qquad \beta=-(\xr R+\rho^2),
\end{equation}
where $\tilde{\Pi}$ denotes the odd extension of Sec.~\ref{sec:oddext} applied to $\Pi(\hD(\psi),\nu,\hE)$ as a function of $\psi$, i.e.
\begin{equation*}
\tilde\Pi(\psi)=\sgn(\psi)\bigl(\Pi(\hD(|\psi|),\nu,\hE)-\Pi(1,\nu,\hE)\bigr).
\end{equation*}
As $0<\nu_+<1$ and $\nu_-<0$, both integrals are real and regular. On the cylinder axis, $\xr\to0$, Eq.~\eqref{eq:OmS} reduces to the solid angle of a cone segment, $\Omega_S=(\theta_2-\theta_1)\bigl(1-|\Dz|/\hA_0\bigr)$, and for a full disk, $\psi_2-\psi_1=2\pi$, it reduces to the result of Ref.~\citep{Paxton_1959}. Thus
\begin{equation}
\mathcal{O}(z')=\sgn(\Dz)\bigl(\Omega_S-\Omega_{T_1}-\Omega_{T_2}\bigr),
\label{eq:Oclosed}
\end{equation}
and with Eq.~\eqref{eq:Ncaps} also $N_{zz}$ is given in closed form.

\subsection{Pieces in the other quadrants}
\label{sec:quadrants}
A piece whose arc lies in the second, third or fourth quadrant is the mirror image of a first-quadrant piece in the $yz$-plane, in both coordinate planes, or in the $xz$-plane, respectively. Let
\begin{equation}
\begin{aligned}
s_x&=\sgn(\cos\theta_0),\qquad s_y=\sgn(\sin\theta_0),\\
\mathsf{D}&=\operatorname{diag}(s_x,s_y,1),
\end{aligned}
\end{equation}
and let $\theta_0^\mathrm{I}=\atantwo(s_y\sin\theta_0,\,s_x\cos\theta_0)\in(0,\pi/2)$ be the centre angle of the reflected piece, whose extent $\delta\theta$, $z_0$ and $\delta z$ are unchanged. Since the field is a polar vector and the tensor transforms accordingly under the reflection $\mathsf{D}$,
\begin{equation}
\bfN(\bfr;\theta_0)=\mathsf{D}\,\bfN^\mathrm{I}(\mathsf{D}\bfr;\theta_0^\mathrm{I})\,\mathsf{D},
\label{eq:reflect}
\end{equation}
where $\bfN^\mathrm{I}$ is the first-quadrant tensor of Secs.~\ref{sec:arc} to \ref{sec:caps}, evaluated at the reflected field point $(s_xx,\,s_yy,\,z)$. In components, Eq.~\eqref{eq:reflect} changes the sign of $N_{xy}$, $N_{yx}$, $N_{xz}$ and $N_{zx}$ in the second quadrant, of $N_{xz}$, $N_{zx}$, $N_{yz}$ and $N_{zy}$ in the third, and of $N_{xy}$, $N_{yx}$, $N_{yz}$ and $N_{zy}$ in the fourth, while the diagonal components are unchanged.

\subsection{The complementary piece}
\label{sec:inverted}
The region between the arc and the corner $3'=(x_2,y_2)$, shown dashed in Fig.~\ref{fig:geometry}, is the complement of the circular piece within the rectangle $x_1\le x'\le x_2$, $y_1\le y'\le y_2$. Its cross section is bounded by the same arc, now with the inward normal $-\rhat$, by the vertical chord at $x'=x_2$ from $y_1$ to $y_2$ with the normal $+\xhat$, and by the horizontal chord at $y'=y_2$ from $x_1$ to $x_2$ with the normal $+\yhat$; for the end surfaces the integration over $x'$ runs from the arc to $x_2$ instead of from $x_1$ to the arc. Repeating the derivation, the tensor of the complementary piece is obtained from Eqs.~\eqref{eq:NIII}, \eqref{eq:chords} and \eqref{eq:QM}--\eqref{eq:Ncaps} by replacing the fixed chord coordinates $x_1\to x_2$ and $y_1\to y_2$ wherever they appear as an argument of $D$ or in the logarithm and arctangent terms, while keeping all integration ranges $[x_1,x_2]$, $[y_1,y_2]$ and $[z_1,z_2]$ and the integrals along the arc unchanged, and by reversing the sign of every contribution, including the arc contribution of Eq.~\eqref{eq:NIII}. Other quadrants are again handled by Eq.~\eqref{eq:reflect}. The two pieces together fill the rectangular prism $[x_1,x_2]\times[y_1,y_2]\times[z_1,z_2]$, and we have verified numerically that the sum of their tensors equals the tensor of that prism \citep{Smith_2010} to machine precision, see Sec.~\ref{sec:verification}.

\section{Verification}
\label{sec:verification}
\begin{figure}
\centering
\includegraphics[width=0.5\textwidth]{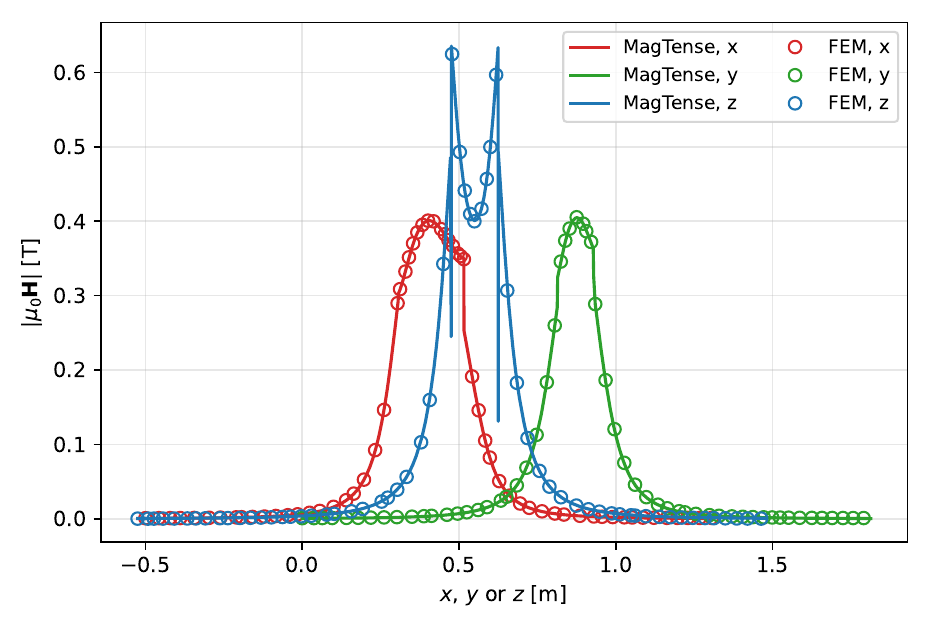}
\caption{The norm of the magnetic field as calculated using the components of the tensor given above in MagTense and as calculated using the FEM framework Comsol, for the circular piece geometry specified in the text, along each of the Cartesian axes through the point $(x,y,z)=(0.406,0.863,0.55)$~m.}
\label{fig:comsol}
\end{figure}

The expressions derived above are verified against a finite element method computation of the magnetic field from a circular piece, and, with the same geometry, against direct numerical integration of Eq.~\eqref{eq:Ndef}. We consider a circular piece that has the following geometry: $R=0.7$~m, $\theta_0=\pi/3$, $\delta\theta=\pi/7$, $z_0=0.35$~m and $\delta z=0.15$~m, and the cylinder axis is displaced from the origin to $(x_c,y_c,z_c)=(0.1,0.3,0.2)$~m. The piece spans $r_0\pm\delta r/2$ with $R=r_0+\delta r/2$, this corresponds to $r_0=0.45$~m and $\delta r=0.5$~m. The magnetization is specified by the vector $\mu_0\bfM=[-0.371,\ -0.270,\ 1.109]$~T, i.e.\ a magnitude of $\mu_0M=1.2$~T along the direction $(-0.310,-0.225,0.924)$. The magnetic field is computed using the finite element framework Comsol \citep{Comsol}, which solves the magnetic scalar potential equation
\begin{equation}
-\nabla\cdot(\mu_0\nabla\phi_M)=-\nabla\cdot(\mu_0\bfM)
\end{equation}
for the given magnetization, after which the magnetic field is calculated as $\bfH=-\nabla\phi_M$. For the FEM model a highly refined finite element mesh is used to ensure a high precision in the computed field, and a sufficiently large volume is modelled to make the boundary conditions not influence the generated magnetic field. The norm of the magnetic field is calculated along each of the Cartesian axes through the point $(x,y,z)=(0.406,0.863,0.55)$~m inside the piece and is shown in Fig.~\ref{fig:comsol}. As can be seen from the figure, there is an excellent agreement between the model framework presented here and the FEM model, also across the surfaces of the piece, where the norm of the field is discontinuous. The relative integrated difference between the two curves is 0.8\,\%, 1.1\,\% and 0.9\,\% along the $x$-, $y$- and $z$-axis, respectively, which is set by the resolution of the finite element mesh at the surfaces of the piece. This example is part of the verification examples for MagTense and is available online \citep{MagTense}.

For the same geometry, the tensor was also compared with a direct adaptive two-dimensional quadrature of Eq.~\eqref{eq:Ndef} over the five surfaces of the piece, at 27 points on the three lines of Fig.~\ref{fig:comsol}, of which four lie inside the piece. The tensor assembled from the expressions of Sec.~\ref{sec:evaluating} agrees with the reference quadrature to $7\times10^{-14}$, which is the precision of the latter.

\section{Conclusion}
We have calculated the magnetic field, formulated as a tensor, for a homogeneously magnetized circular piece, i.e.\ the part of a finite cylinder cut off by two planes parallel to the coordinate axes, and for its complement. All nine components of the tensor are given in closed form in terms of elementary functions and incomplete elliptic integrals of the first, second and third kind. There is an excellent agreement between the magnetic field of a circular piece calculated using the tensor, implemented in MagTense, and the field computed using direct numerical integration and a finite element approach.

\section*{Data statement}
The derived analytical expressions for the demagnetization tensor of the circular piece and of its complement are available through the MagTense framework \citep{MagTense}, available at \url{www.magtense.org}, as is the verification example shown in Fig.~\ref{fig:comsol}.

\section*{Declaration of generative AI and AI-assisted technologies in the manuscript preparation process}
During the preparation of this work the authors used Anthropic Claude, model Fable 5.1 to suggest analytical solutions to some of the end surface integrals, as well as draft text on these. The model was also used to more quickly reformat equations, substitute expressions and simplify the mathematical expressions. After using this model, the authors reviewed and edited the content as needed and takes full responsibility for the content of the published article.

\section*{Acknowledgements}
This work was supported by the Carlsberg Foundation Semper Ardens Advance project CF24-0920 entitled ``Novel magnets through interdisciplinarity and nanocomposites'' and the Independent Research Fund Denmark grant ``The next generation permanent magnet for the green transition'', 4307-00105B.

\bibliographystyle{IEEEtran}

\end{document}